\documentclass[pra,twocolumn,nofootinbib,eqsecnum,floatfix,showpacs]{revtex4}
\usepackage{bm} 
\usepackage{amsmath}

\begin{document}

\title{{\bf Insufficiency of the Quantum State
for Deducing Observational Probabilities}
\thanks{Alberta-Thy-17-08, arXiv:0808.0722 [hep-th]}}

\author{Don N. Page}
\email{don@phys.ualberta.ca}

\affiliation{Theoretical Physics Institute\\
Department of Physics, University of Alberta\\
Room 238 CEB, 11322 -- 89 Avenue\\
Edmonton, Alberta, Canada T6G 2G7}

\date{2008 September 17}

\begin{abstract}

It is usually assumed that the quantum state is sufficient for deducing all
probabilities for a system.  This may be true when there is a single observer,
but it is not true in a universe large enough that there are many copies of an
observer.  Then the probability of an observation cannot be deduced simply from
the quantum state (say as the expectation value of the projection operator for
the observation, as in traditional quantum theory).  One needs additional rules
to get the probabilities.  What these rules are is not logically deducible from
the quantum state, so the quantum state itself is insufficient for deducing
observational probabilities.  This is the measure problem of cosmology.

\end{abstract}


\maketitle

\section{Introduction}

All probabilities for a system are believed to be encoded in its quantum state. 
This may be true, but there is the question of how to decode the quantum state
to give these probabilities.  In traditional quantum theory, the probabilities
are given by the expectation values of projection operators.  Once a possible
observation is specified (including the corresponding projection operator), then
its probability is given purely by the quantum state as the expectation value
the state assigns to the projection operator, a mathematization of the Born rule
\cite{Born}.

This prescription works well in ordinary single laboratory settings, where there
are no copies of the observer.  Then distinct observations are mutually
exclusive, so that different ones cannot both be observed.  If one assigns a
projection operator to each possible distinct observation in a complete
exhaustive set, then these projection operators will be orthonormal, and their
(nonnegative) expectation values will sum to unity, which are conditions
necessary for them to be interpreted as the probabilities of the different
possible observations.

However, in cosmology there is the possibility that the universe is so large
that there are many copies of each observer, no matter how precisely the
observer is defined.  This raises the problem \cite{HS,typdef,typder,cmwvw} that
two observations that are seen as distinct for an observer are not mutually
exclusive in a global viewpoint; both can occur for different copies of the
observer (though neither copy may be aware of that).  This would not be a
problem for a putative superobserver who can observe all possible sets of
observations by all observers over the entire universe, but it is a problem for
the assignment of normalized probabilities for the possible observations that
are distinct for each copy of the observer.  The result \cite{typder} is that
one cannot get such a set of normalized probabilities as the expectation values
of projection operators in the full quantum state of the universe.

One can still postulate that there are rules for getting the probabilities of
all possible observations from the quantum state, but then the question arises
as to what these rules are.  Below we shall give examples of several different
possibilities for these rules, showing that they are not uniquely determined and
thus that they are logically independent of the question of what the quantum
state is.  Therefore, the quantum state just by itself is insufficient to
determine the probabilities of observations.

The main application of the logical independence of the probability rules is to
the {\it measure problem} in cosmology (see \cite{cmwvw} for many references),
the problem of how to make statistical predictions for observations in a
universe that may be so large that almost all theoretically possible
observations actually occur somewhere.  The logical independence implies that
the solution to the measure problem is not just the quantum state of the
universe but also other independent elements, the rules for getting the
probabilities of the observations from the quantum state.

\section{Observational probabilities with many copies of the observer}

A goal of science is to come up with theories $T_i$ that predict the
probabilities of results of observations (observational results).  Here for
simplicity I shall assume that there is a countable set of possible distinct
observations $O_j$ out of some exhaustive set of all such observations.  This
set of possible observations might be, for example, all possible conscious
perceptions \cite{SQM}, all possible data sets for one person, all possible
contents for an eprint arXiv, or all possible data sets for a human scientific
information gathering and utilizing system \cite{HS}.  If one imagines a
continuum for the set of observations (which seems to be logically possible,
though not required), in that case I shall assume that they are binned into a
countable number of exclusive and exhaustive subsets that each may be considered
to form one distinct observation $O_j$.  Then the goal is to calculate the
probability $P_j(i) \equiv P(O_j|T_i)$ for the observation $O_j$, given the
theory $T_i$.

One might think that once one has the quantum state, there would be a standard
answer to the question of the probabilities for the various possible
observations.  For example \cite{typder}, one might take traditional quantum
theory (what I there called standard quantum theory) to give the probability
$P_j(i)$ of the observation as the expectation value, in the quantum state given
by the theory $T_i$, of a projection operator $\mathbf{P}_j$ onto the
observational result $O_j$.  That is, one might take
\begin{equation}
P_j(i) = \langle \mathbf{P}_j \rangle_i, 
\label{standard}
\end{equation}
where $\langle \rangle_i$ denotes the quantum expectation value of whatever is
inside the angular brackets in the quantum state $i$ given by the theory $T_i$. 
This traditional approach works in the case of a single laboratory setting where
the projection operators onto different observational results are orthogonal,
$\mathbf{P}_j \mathbf{P}_k = \delta_{jk}\mathbf{P}_j$ (no sum over repeated
indices).

However \cite{typder,cmwvw}, in the case of a sufficiently large universe, one
may have observation $O_j$ occurring `here' and observation $O_k$ occurring
`there' in a compatible way, so that $\mathbf{P}_j$ and $\mathbf{P}_k$ are not
orthogonal.  Then the traditional quantum probabilities given by Eq.
(\ref{standard}) will not be normalized to obey
\begin{equation}
\sum_j P_j(i) = 1  . 
\label{norm}
\end{equation}
Thus one needs a different formula for normalizable probabilities of a mutually
exclusive and exhaustive set of possible observations, when distinct
observations within the complete set cannot be described by orthogonal
projection operators.

Although many other rules are also possible, as I shall illustrate below, the
simplest class of modifications of Eq. (\ref{standard}) would seem to be to
replace the projection operators $\mathbf{P}_j$ with some other {\it observation
operators} $\mathbf{Q}_j(i)$ normalized so that $\sum_j \langle \mathbf{Q}_j(i)
\rangle_i = 1$, giving
\begin{equation}
P_j(i) = \langle \mathbf{Q}_j(i) \rangle_i . 
\label{eventual}
\end{equation}
Of course, one also wants $P_j(i) \geq 0$ for each $i$ and $j$, so one needs
to impose the requirement that the expectation value of each observation
operator $\mathbf{Q}_j(i)$ in each theory $T_i$ is nonnegative.

The main point \cite{typder,cmwvw} is that in cases with more than one copy of
the observer, such as in a large enough universe, one cannot simply use the
expectation values of projection operators as the probabilities of observations,
so that, if Eq. (\ref{eventual}) is to apply, each theory must assign a set of
observation operators $\mathbf{Q}_j(i)$, corresponding to the set of possible
observations $O_j$, whose expectation values are used instead as the
probabilities of the observations.  Since these observation operators are not
given directly by the formalism of traditional quantum theory, they must be
added to that formalism by each particular complete theory.  In other words, a
complete theory $T_i$ cannot be given merely by the dynamical equations and
initial conditions (the quantum state), but it also requires the set of
observation operators $\mathbf{Q}_j(i)$ whose expectation values are the
probabilities of the observations $O_j$ in the complete set of possible
observations (or else some other rule for the probabilities, if they are not to
be expectation values of operators).  The probabilities are not given purely by
the quantum state but have their own logical independence in a complete theory.

Let us suppose that we can hypothetically partition spacetime into a countable
set of disjoint regions labeled by the index $L$, with each region having its
own reference frame and being sufficiently small that for each $L$ separately
there is a set of orthogonal projection operators $\mathbf{P}_j^L$ whose
expectation values give good approximations to the probabilities that the
observations $O_j$ occur within the region $L$.  Each region has its own algebra
of quantum operators, and for simplicity I shall make the somewhat unrealistic
assumption that the regions are either spacelike separated or are so far apart
that each operator in one region, such as $\mathbf{P}_j^L$, commutes with each
observable in a different region, such as $\mathbf{P}_k^M$ for $L\neq M$. 
(However, I am assuming that each observation $O_j$ can in principle occur
within any of the regions, so that the content of the observation is not
sufficient to distinguish what $L$ is; the observation does not determine where
one is in spacetime.  One might imagine that an observation determines as much
as it is possible to know about some local region, but it does not determine the
properties outside, which might go into the specification of the index $L$ that
is only known to a hypothetical superobserver that makes the partition.)

Now one might propose that one construct the projection operator
\begin{equation}
\mathbf{P}_j = \mathbf{I} - \prod_L (\mathbf{I} - \mathbf{P}_j^L) 
\label{existence}
\end{equation}
(this being the only place where I need the $\mathbf{P}_j^L$'s to commute for
different $L$) and use it in Eq. (\ref{standard}) to get a putative probability
of the observation $O_j$ in the quantum state given by the theory $T_i$. 
Indeed, this is essentially in quantum language \cite{typder} what Hartle and
Srednicki \cite{HS} propose, that the probability of an observation is the
probability that it occur at least somewhere.  However, because the different
$\mathbf{P}_j$'s defined this way are not orthogonal, the resulting traditional
quantum probabilities given by Eq. (\ref{standard}) will not be normalized to
obey Eq. (\ref{norm}).  This lack of normalization is a consequence of the fact
that even though it is assumed that two different observations $O_j$ and $O_k$
(with $j \neq k$) cannot both occur within the same region $L$, one can have
$O_j$ occurring within one region and $O_k$ occurring within another region. 
Therefore, the existence of the observation $O_j$ at least somewhere is not
incompatible with the existence of the distinct observation $O_k$ somewhere
else, so the sum of the existence probabilities is not constrained to be unity.

If one were the hypothetical superobserver who has access to what is going on in
all the regions, one could make up a mutually exclusive and exhaustive set of
joint observations occurring within all of the regions.  However, for us
observers who are confined to just one region, the probabilities that such a
superobserver might deduce for the various combinations of joint observations
are inaccessible for us to test or to use to predict what we might be expected
to see.  Instead, we would like probabilities for the observations we ourselves
can make.  I am assuming that each $O_j$ is an observational result that in
principle we could have, but that we do not have access to knowing which region
$L$ we are in.  (The only properties of $L$ that we can know are its local
properties that are known in the observation $O_j$ itself, but that is not
sufficient to determine $L$, which might be determined by properties of the
spacetime beyond our local knowledge.)

\section{Examples of different observational probabilities for the same quantum
state}

Let us demonstrate the logical freedom in the rules for the observational
probabilities $P_j(i) \equiv P(O_j|T_i)$ by exhibiting various examples of what
they might be. For simplicity, let us restrict attention to theories $T_i$ that
all give the same pure quantum state $|\psi\rangle$, which can be written as a
superposition, with fixed complex coefficients $a_N$, of component states
$|\psi_N\rangle$ that each have different numbers $N$ of observational regions:
\begin{equation}
|\psi\rangle = \sum_{N=0}^{\infty} a_N |\psi_N\rangle, 
\label{state}
\end{equation}
where $\langle\psi_M|\psi_N\rangle = \delta_{MN}$.  (Different values of $N$
model different sizes of universes produced by differing amounts of inflation in
the cosmological measure problem.) The different theories will then differ only
in the prescriptions they give for calculating the observational probabilities
$P_j(i)$ from the single quantum state $|\psi\rangle$.  These differences will
illustrate the logical independence of the observational probabilities from the
quantum state, the fact that the observational probabilities are not uniquely
determined by the state.

In each component state, the index $L$ can run from 1 to $N$ (except for the
component state $|\psi_0\rangle$, which has no observational regions at all). 
As above, let us suppose that $\mathbf{P}_j^L$ is a complete set of orthogonal
projection operators for the observation $O_j$ to occur in the region $L$.  Then
if only the region $L$ existed, and the state were $|\psi_N\rangle$, then the
quantum probability of the observation $O_j$ would be
\begin{equation}
p_{NLj} = \langle\psi_N|\mathbf{P}_j^L|\psi_N\rangle. 
\label{Lprob}
\end{equation}
However, in reality, even just in the component state $|\psi_N\rangle$ for
$N>1$, there are other regions where the observation could occur, so the total
probability $P_j(i)$ for the observation $O_j$ in the theory $T_i$ can be some
$i$-dependent function of all the $p_{NLj}$'s.  The freedom of this function is
part of the independence of the observational probabilities from the quantum
state itself.

Let us define existence probabilities $p_j(i)$ that might not be normalized to
add up to unity when summed over $j$ for the different possible observational
results $O_j$, and then use $P_j(i)$ for normalized observational probabilities
obeying Eq. (\ref{norm}).  The different indices $i$ will denote different
theories, in this case different rules for calculating the probabilities, since
for simplicity we are assuming that all the theories have the same quantum state
$|\psi\rangle$.

Next, let us turn to different possible examples.

For theory $T_1$, let us suppose that the existence probability $p_j(1)=0$ if
there is no region $L$ in any nonzero component of the quantum state
($|\psi_N\rangle$ with $a_N \neq 0$) that has a positive expectation value for
$\mathbf{P}_j^L$, so that $\sum_{N,L} |a_N|^2 p_{NLj} = 0$, but that $p_j(1)=1$
otherwise, that is if $\sum_{N,L} |a_N|^2 p_{NLj} > 0$.  This theory is
essentially taking the Everett many worlds interpretation to imply that if there
is any nonzero amplitude for the observation to occur, it definitely exists
somewhere in the many worlds (and hence has existence probability unity).

Now of course these existence probabilities $p_j(1)$ are not necessarily
normalized, since $M>1$ of them can be unity, with the rest zero, so the sum of
these existence probabilities is $M$.  However, one could say that so far as the
probability goes of making a particular one of the $M$ actually existing
observations, that could be considered equally divided between the $M$ actually
existing possibilities (if $M>0$), so that one has normalized observational
probabilities $P_j(1)=p_j(1)/M$.  This would be the theory that every
observation that actually does exist is equally probable.

For theory $T_2$, define the existence probabilities to be $p_j(2) =
\langle\psi|\mathbf{P}_j|\psi\rangle$, the expectation value in the full quantum
state $|\psi\rangle$ of the projection operator $\mathbf{P}_j$ defined by Eq.
(\ref{existence}) for the existence of the observation $O_j$ in at least one
region $L$.  This is not the full many-worlds existence probability, which is
unity if the observation does occur somewhere, but it might be regarded as the
quantum probability for a superobserver to find that at least one instance of
the observation $O_j$ occurs.

Again, these existence probabilities will not in general sum to unity, but one
can normalize by dividing by the sum $M$ (now generically not an integer) to get
normalized probabilities $P_j(2)=p_j(2)/M$.  For a universe that is very large
(e.g., most of the $|a_N|^2$'s concentrated on very large $N$'s), one would
expect a large number of $p_j(2)$'s to be very near unity (since it would be
almost certain that the observation $O_j$ occurs at least somewhere among the
huge number of regions), so that there will be a large number of nearly equal
but very small $P_j(2)$'s.

For theory $T_3$, refrain from defining the existence probabilities $p_j(3)$
(since I am regarding it sufficient for a theory to prescribe only the
observational probabilities of observations in regions within the universe, not
for observations by some putative superobserver).  Instead, define unnormalized
observational measures
\begin{eqnarray}
\mu_j(3) = \sum_L \langle\psi|\mathbf{P}_j^L|\psi\rangle
         = \sum_{N=1}^\infty\sum_{L=1}^N|a_N|^2 p_{NLj}.
\label{3measure}
\end{eqnarray}
Then normalize these to let the observational probabilities be defined as
\begin{eqnarray}
P_j(3) = \frac{\mu_j(3)}{\sum_k \mu_k(3)}.
\label{3prob}
\end{eqnarray}

For theory $T_4$, define unnormalized observational measures
\begin{eqnarray}
\mu_j(4) = \sum_{N=1}^\infty\frac{1}{N}\sum_{L=1}^N|a_N|^2 p_{NLj}
\label{4measure}
\end{eqnarray}
and then normalized observational probabilities
\begin{eqnarray}
P_j(4) = \frac{\mu_j(4)}{\sum_k \mu_k(4)}.
\label{4prob}
\end{eqnarray}

Theory $T_3$ in its sum over $L$ effectively weights each component state
$|\psi_N\rangle$ by the number of observational regions where the observation
$O_j$ can potentially occur.  On the other hand, theory $T_4$ has an average
over $L$ for each total number $N$ of observational regions, so that component
states $|\psi_N\rangle$ do not tend to dominate the probabilities for
observations just because of the greater number of observational opportunities
within them.  Theory $T_3$ is analogous to volume weighting in the cosmological
measure, and theory $T_4$ is analogous to volume averaging \cite{cmwvw}.

Of these four rules, $T_1$ does not give the probabilities as the expectation
values of natural observation operators $\mathbf{Q}_j(i)$, but the other three
do, with the corresponding observation operators being
\begin{eqnarray}
\mathbf{Q}_j(2) &=& \frac{\mathbf{P}_j}{\langle\sum_k\mathbf{P}_k\rangle_i},
\nonumber \\ 
\mathbf{Q}_j(3) &=& \frac{\sum_L\mathbf{P}_j^L}
{\langle\sum_k\sum_L\mathbf{P}_k^L\rangle_i},
\nonumber \\  
\mathbf{Q}_j(4) 
&=& \frac{\sum_{N=1}^\infty\frac{1}{N}
            \sum_{L=1}^N\mathbf{P}_N\mathbf{P}_j^L\mathbf{P}_N}
         {\langle\sum_k\sum_{N=1}^\infty\frac{1}{N}
            \sum_{L=1}^N\mathbf{P}_N\mathbf{P}_k^L\mathbf{P}_N\rangle_i},
\label{obsop}
\end{eqnarray}
where $\mathbf{P}_N = |\psi_N\rangle\langle\psi_N|$ is the projection operator
onto the component state with $N$ observation regions.

These examples show that there is not just one unique rule for getting
observational probabilities from the quantum state.  It remains to be seen what
the correct rule is.  Of the four examples given above, I suspect that with a
suitable quantum state, theory $T_4$ would have the highest likelihood $P_j(i)$,
given our actual observations, since theories $T_1$ and $T_2$ would have the
normalized probabilities nearly evenly distributed over a huge number of
possible observations, and theory $T_3$ seems to be plagued by the Boltzmann
brain problem \cite{cmwvw}.  One might conjecture that theory $T_4$ can be
implemented in quantum cosmology to fit observations better than other
alternatives \cite{cmwvw}.

Thus we see that in a universe with the possibility of multiple copies of an
observer, observational probabilities are not given purely by the quantum state,
but also by a rule to get them from the state.  There is logical freedom in what
this rule is (or in what the observation operators $\mathbf{Q}_j(i)$ are
if the rule is that the probabilities are the expectation values of these
operators).  In cosmology, finding the correct rule is the measure problem.

\section*{Acknowledgments}

I am grateful for discussions with Tom Banks, Raphael Bousso, Sean Carroll,
Brandon Carter, Alan Guth, James Hartle, Andrei Linde, Seth Lloyd, Juan
Maldacena, Mark Srednicki, Alex Vilenkin, an anonymous referee, and others, and
especially for a long email debate with Hartle and Srednicki over typicality
that led me to become convinced that there is logical freedom in the rules for
getting observational probabilities from the quantum state.  This research was
supported in part by the Natural Sciences and Engineering Research Council of
Canada.

\end{document}